\documentclass[a4paper,12pt]{article}%
\usepackage{amsfonts}
\usepackage{makeidx}
\usepackage{latexsym}
\usepackage{amsmath}
\usepackage{color}
\usepackage{amssymb}
\usepackage{graphicx, amsmath, amsthm, amssymb, bm, boxedminipage}
\usepackage{graphicx}%
\allowdisplaybreaks[4]
\allowdisplaybreaks[4]

\numberwithin{equation}{section}
\theoremstyle{plain}
\newtheorem{thm}{Theorem}[section]

\begin{document}

\title{Identifying Heterogeneous Decision Rules From Choices When Menus Are
Unobserved\thanks{Department of Economics, McGill University,
larry.epstein@mcgill.ca and kaushil.patel@mail.mcgill.ca. We are extremely
indebted to Rohan Dutta who was instrumental behind the scenes from conception
through execution. We thank also participants at the McGill applied micro
breakfast for their suggestions. }}
\author{Larry G. Epstein
\and Kaushil Patel}
\maketitle
\date{}

\begin{abstract}
Consider aggregate choice data from a population with heterogeneity in both
preferences (or more general decision rules) and in menus, and where the
analyst has limited information about how menus are distributed across the
population. We determine what can be inferred from aggregate data about the
distribution of preferences by identifying the set of all distributions that
are consistent with the preceding. Our main theorem strengthens and
generalizes existing results on such identification and provides an
alternative analytical approach (using capacities) to study the problem. We
show further that our model and results are applicable, after suitable
reinterpretation, to other contexts. One such application is to the robust
identification of the distribution of updating rules from the observed
population distribution of beliefs, while respecting that differences in
information are unobserved and poorly understood.

\medskip

\bigskip


\end{abstract}

\newpage

\section{Introduction}

\subsection{Motivation and outline}

Consider the problem of explaining the distribution of choices in a
heterogeneous population. Denote by $\lambda$ the probability distribution of
chosen alternatives, the data. A common approach is to posit heterogeneity in
decision rules (or underlying preferences) and possibly also in the menus from
which alternatives are chosen. A decision rule $d$ specifies the alternative
$d\left(  A\right)  $ chosen from each menu $A$; the set of all decision rules
is $\mathcal{D}$. An individual with decision rule $d$ faces menu $A$ with
probability $\pi_{d}\left(  A\right)  $. Decision rules are distributed
according to a probability measure $Q$ that is to be inferred from the data,
while the collection of probability measures $\{\pi_{d}\}_{d\in\mathcal{D}}$
is known to the analyst (possibly up to unknown
parameters).\footnote{Filiz-Ozbay and Masatlioglu (2023) call this a
random-choice model (RCM), defined by a probability distribution over a
collection of choice functions (potentially irrational). They axiomatize a
specific class of RCMs under the assumption of rich stochastic choice data.}
Accordingly, she seeks $Q$ satisfying, \emph{for the given }$\{\pi_{d}%
\}_{d\in\mathcal{D}}$,
\begin{equation}
\lambda\left(  a\right)  =%
{\displaystyle\sum\limits_{d}}
{\displaystyle\sum\limits_{A}}
Q\left(  d\right)  \pi_{d}(A)~\boldsymbol{1}_{d\left(  A\right)  =a}\text{, }
\label{Rat0}%
\end{equation}
for all alternatives $a$. Then empirical frequencies are rationalized by the
heterogeneity in decision rules described by $Q$. Of particular interest is
the set of all rationalizing $Q$s (the sharp identified region).

The above model is general in that it covers the bulk of the discrete choice
literature where various special cases are adopted;\footnote{We are ignoring
covariates that often appear in this literature, and that could be added
below, because they are not germane to our contribution. We adopt a
streamlined formulation in order to maximize transparency of the theoretical
point of this paper.} for example, the traditional assumption (McFadden 1974)
that the menu corresponding to each choice is observed corresponds to the
special case where $\pi_{d}(A)=1$ for some $A$. However, data about menus that
would support knowledge of the conditional probabilities $\pi_{d}$ are often
unavailable (see Manski (1977) and the overviews and many references in
Barseghyan et al (2021, pp. 2016-2017, 2041-2043) and Azrieli and Rehbeck
(2023)). Notably, decision models based on consideration sets (Abaluck and
Adams-Prassl 2021, Cattaneo et al 2020, Manzini and Mariotti 2014, Masatlioglu
et al 2012) or rational inattention (Caplin et al 2019) view choices as made
from subjective menus, thus arguing against their observability.\footnote{To
be clear, we use "menu" to refer to the set from which an alternative is
chosen by maximizing preference or by applying another decision rule.
Consequently, it may be a strict subset of the objective feasible set, (for
example, a consideration set), that is determined by the individual's
cognitive deliberation process and is unobservable to the analyst.} One is led
to the concern that conclusions about the identified set of measures $Q$ that
are based on (\ref{Rat0}) sometimes rely on ad hoc assumptions about menus.

An objective in this paper is to robustify the above model by incorporating
the analyst's imperfect knowledge about menus. One alternative to the perfect
information assumption is complete ignorance about menus - "anything goes" for
specifications of $\pi_{d}$s. However, in general, one would expect there to
be partial information about the menu process. Therefore, we admit a range of
assumptions about the analyst's information that are intermediate between
complete ignorance and perfect knowledge. In all cases, we show (Theorem
\ref{thm-main}) that the implied sharp identified set of distributions
consists of all measures $Q$ satisfying a finite set of linear inequalities
and hence forms a polytope (a convex set with finitely many extreme points);
in particular, it is computationally tractable.

We adopt a novel formulation using convex (or supermodular) capacities and
their cores. (The appendix collects the few basic definitions and facts
regarding capacities that are used below; a very accessible and comprehensive
reference is Grabisch (2016).) Capacities are set functions that generalize
probability measures in order to permit a role in the representation of
beliefs for limited information and the resulting limited confidence in any
single probability measure - in other terms, uncertainty about probabilities.
They arise in decision theory, notably in Schmeidler's (1989) Choquet expected
utility theory, where convexity of the capacity is identified with aversion to
such uncertainty and where convexity characterizes the Choquet models that
conform also to multiple-priors utility (Gilboa and Schmeidler
1989).\footnote{Convex capacities, or equivalently their conjugates, known as
2-alternating, are important also in statistical theory (in proving an
extension of the Neyman-Pearson Lemma (Huber and Strassen 1973) and in
supporting a version of Bayes' theorem for capacities (Wasserman and Kadane
1990). They appear also in cooperative game theory as characteristic
functions. However, the epistemic interpretation is a better fit here.} For
our purposes, the key technical feature of convex capacities is that "the core
of a mixture of capacities equals the mixture of their cores" (see
(\ref{mixturelinear}) for a formal statement). Given our formulation, this
property leads to a short transparent (indeed elementary) proof of our theorem
that applies to and unifies all of our specifications. We view this simplicity
and the associated epistemic perspective as a strength and a contribution.

The scope of our results merits emphasis. Thus far we have interpreted the
paper as addressing heterogeneity in choice assuming heterogeneity in decision
rules and the unobservability of menus. However, with suitable
reinterpretation of the symbols in the formal model, Theorem \ref{thm-main}
applies also to other contexts where one seeks the identification of
heterogeneity that is robust to unobservables. A concrete setting where we
describe such an application is to the rationalization of the distribution of
beliefs in a population (see section \ref{section-updating}).\footnote{Two
other settings are outlined in the concluding section \ref{section-conclude}.}
Individuals often disagree about the likelihoods of future events. Two
candidate reasons are differences in information and in updating, both of
which are often unobservable to the analyst (outside the laboratory).
Unobservability of information is a common assumption. However, it is
typically accompanied by the assumption that all updating is Bayesian,
contrary to the abundant experimental evidence of a number of systematic
biases (see the seminal study by Tversky and Kahneman (1974), and the many
references in Camerer (1995), Rabin (1998)); particularly relevant for the
sequel is evidence of under/overreaction (see Benjamin (2019) and Ba, Bohren
and Imas (2023) and the many references therein). Such evidence motivates our
hypothesis that updating rules may differ between individuals, with universal
Bayesian updating being a very special case. We admit a family of updating
rules that deviate from Bayesian updating by either overreacting or
underreacting to information and we (partially) identify the population
distribution of updating rules, while respecting that little may be known
about differences in information. We also provide conditions under which any
rationalization of the distribution of beliefs exhibits underreaction (or
overreaction, respectively) "on average."

Our main theorem, Theorem \ref{thm-main}, is, in fact, equivalent to Theorem 4
of Strassen (1965) when specialized as here so that all sets are finite. An
important difference is our drastically simpler elementary proof. We view the
simpler proof as significant not as a mathematical contribution, but rather
because it enhances transparency and accessibility of the theorem which, we
believe, may help to expose and promote it as a useful tool for economists.
The other value-added over Strassen is our demonstration of the theorem's
usefulness as outlined above.

\subsection{Related literature}

First we relate our contribution to some recent papers in discrete choice (and
related econometrics) that also weaken a priori assumptions about menus.
Barseghyan et al (2021) study identification in a random utility model where
the distribution of menus in the population is unknown. Two differences from
our model are that: they assume preference maximization (particularly, Sen's
$\alpha$ condition) rather than general decision rules, and they assume that
all menus of size at least $\kappa$, ($\kappa\geq2$), a parameter specified by
the analyst, are conceivable for any individual conditional on her preference
order, while we allow the set of conceivable menus to be arbitrary. Moreover,
they deal \emph{only} \emph{with the case of complete ignorance} of the menu
process, for which their characterization of the sharp identified set
corresponds (apart from their inclusion of covariates) to our
complete-ignorance result in Theorem \ref{thm-main}.\footnote{Minor
differences are described following the statement of our theorem in section
\ref{section-thm}.} Their proof also differs from ours in that it applies the
theory of random sets. The significance of this difference is that random sets
induce a belief function which is a very special kind of convex capacity that
precludes many of the richer information structures (those short of complete
ignorance about menus) that are accommodated in our theorem. \ Lu (2022)
assumes that all conceivable menus are bounded above and below in the sense of
set inclusion, and that the bounding menus are \emph{known}. He uses the
latter, and the assumption that decision rules satisfy Sen's $\alpha$, to
describe a superset of the identified region. In contrast, our conditions on
$Q$ are both necessary and sufficient for $Q$ to rationalize the data, thus
yielding the sharp identified region. Azrieli and Rehbeck (2023) also study
what can be learned from aggregate choice frequencies, but with several
differences from the present paper. A major difference is that they assume
that the marginal empirical distributions of both menus and choices are known
(constitute the data). In their study of random utility models, they assume
that menus are homogeneous across decision makers, (that is, the distribution
of menus does not depend on the decision rule), while we allow for correlation
between menus and decision rules. An example in section \ref{section-thm}
shows that menu-homogeneity can strictly shrink the sharp identification
region. Regarding proof arguments, they also highlight their use of \ "known
properties of the core," though these do not not include the key property that
we exploit here, and they borrow more from cooperative game theory than from
decision theory and thus do not emphasize epistemics in their interpretations.
Further, their proofs (specifically for their Proposition 9) use not only core
properties but also network flow arguments (based on a version of Hall's
marriage theorem), while we use only the single mixture property of the core
noted above.\footnote{None of the above papers invoke, or cite, Strassen
(1965).}

Dardanoni et al (2020) also explore what can be inferred from aggregate choice
data, though their focus is on cognitive heterogeneity rather than on
preference (or decision rule) heterogeneity. Individuals differ in cognitive
"type" and, given an objective feasible set, they arrive at different
consideration sets (menus in our terminology); further, they do so in a way
that conforms to specific functional forms - the "consideration capacity
model" (which limits the cardinality of the consideration set) or the
"consideration probability model" (Manzini and Mariotti 2014). In the section
most closely related to our paper, where preferences are unobservable and
heterogeneous, they assume that choices are observed from multiple "occasions"
across which both the feasible set and cognitive heterogeneity are stable.
With this rich dataset and functional form restrictions they prove point
identification of the distribution of cognitive types in the consideration
capacity model. Roughly speaking, from the perspective of our formal
framework, they severely restrict the distribution of decision rules and aim
at identification of the menu formation process ($\pi_{d}$), which reflects
the distribution of cognitive type. Unsurprisingly, their proof arguments are
much different than ours.

Doval and Eilat (2023) study the setting where the analyst knows the marginal
over an agent's actions and the prior over states of the world, but does not
know the distribution of actions given realizations of the states of the
world. They ask when two such marginals (over actions and states, similar to
the dataset in Azrieli and Rehbeck 2023) can be rationalized (in the sense of
a Bayes correlated equilibrium) as the outcome of the agent learning something
about the state before taking an action. Their characterization result is two
systems of linear inequalities that are necessary and sufficient for the
dataset to be consistent with a Bayes correlated equilibrium. One of these can
be established using our "mixture of cores" property. Their proof relies on
both Strassen (1965, Theorem 3) and on network flow arguments.

\section{Robust identification\label{section-main}}

\subsection{Preliminaries}

The (finite) universal set of alternatives is $X$, and the set of probability
distributions or measures on $X$ is denoted $\Delta\left(  X\right)  $. Each
individual in a finite population faces a menu, a subset of $X$, from which
she chooses one alternative. The collection of all "relevant" menus is denoted
$\mathcal{A}$, with generic element $A$. The collection $\mathcal{A}$ is a
primitive, determined by the analyst. Another primitive is a finite set
$\mathcal{D}$ of decision rules, where, for each $d$ in $\mathcal{D}$,
$d\left(  A\right)  $ denotes the alternative that $d$ chooses from the menu
$A$ in $\mathcal{A}$. We do not impose any requirements on decision rules, for
example, they need not be derived from preference maximization.

The "data" to be explained are represented by $\lambda\in\Delta\left(
X\right)  $, the empirical frequency distribution of chosen alternatives
across the population.
The analyst's view of the menu formation process determines what constitutes
an "explanation." We assume that she is certain that only menus in
$\mathcal{A}$ are relevant, but otherwise she has limited understanding of how
menus are determined; in particular, she cannot be confident in any single
conditional probability distribution over menus $\pi_{d}\in\Delta\left(
\mathcal{A}\right)  $, which suggests modeling via a \emph{set }of conditional
distributions. We proceed in this way, though with a slight twist as explained next.

Let%
\begin{equation}
C_{d}=\{d\left(  A\right)  :A\in\mathcal{A}\mathbf{\}}\text{.} \label{Di}%
\end{equation}
Thus $C_{d}$ denotes the set of all alternatives that can be chosen by $d$ for
some menu.\ For a given $d$, the analyst can be sure that an element of
$C_{d}$ will be selected, but since the choice depends on the menu, her
limited knowledge of menus affects her view of which choice is associated with
$d$. For any given distribution over menus $\pi_{d}\in\Delta\left(
\mathcal{A}\right)  $, induced beliefs over alternatives are given by
$\rho_{d}\in\Delta\left(  C_{d}\right)  $, where
\begin{equation}
\rho_{d}\left(  a\right)  =\pi_{d}\left(  \{A\in\mathcal{A}:d\left(  A\right)
=a\}\right)  \text{, \ for every }a\in X\text{.} \label{rhoi}%
\end{equation}
Using $\rho_{d}$, (\ref{Rat0}) implies that
\begin{equation}
\lambda\left(  a\right)  =%
{\displaystyle\sum\limits_{d}}
Q\left(  d\right)  \rho_{d}(a)\text{,\ \ \ for every }a\in X\text{,}
\label{Rat0-rho}%
\end{equation}
where menus have been eliminated and distributions over alternatives are
described by both the empirical measure $\lambda$ and by the "explanatory"
measures $\{\rho_{d}\}$. To proceed, we adopt as the benchmark notion of an
explanation of $\lambda$ that "(\ref{Rat0-rho}) is satisfied by the known
$\{\rho_{d}\}$," thus replacing "(\ref{Rat0}) is satisfied by the known
$\{\pi_{d}\}$." In fact, we describe later (section \ref{section-thm}) why the
two benchmarks lead to identical results in the present discrete choice
setting. However, in other contexts, such as in the effort example in section
\ref{section-conclude} there may be no obvious counterpart of (\ref{Rat0}),
while the model based on (\ref{Rat0-rho}) is applicable.

\subsection{Rationalization}

We define what it means for a measure $Q$ over decision rules to rationalize
the empirical measure $\lambda$. In the extreme case where the analyst knows
the distributions over menus this is expressed by (\ref{Rat0-rho}) using the
known conditionals $\{\rho_{d}\}$. One can capture the other extreme of
complete ignorance by requiring that (\ref{Rat0-rho}) is satisfied for
\emph{some} conditionals $\{\rho_{d}\}$, restricting them only to reflect
certainty that $d$ chooses an element in $C_{d}$, that is, $\rho_{d}\in
\Delta\left(  C_{d}\right)  $ for every $d$. The associated robustness may be
desirable but comes with costs (that we formalize below). First, if "anything
goes," then the identified region for any given data $\lambda$ is large.
Second, with such weak maintained assumptions, (almost) every $\lambda$ can be
rationalized by some $Q$. Consequently, and also because there are situations
in which there exists partial information about the menu process, we propose a
model that also accommodates such intermediate situations.

To model the presence of some information, we assume that, for each $d$, only
distributions $\rho_{d}$ that lie in the set $\mathcal{R}_{d}\subset
\Delta\left(  C_{d}\right)  $, determined by the analyst, are deemed relevant.
This leads to the following definition: Say that $Q\in\Delta\left(
\mathcal{D}\right)  $ \emph{rationalizes} $\lambda$ \emph{given}
$\{\mathcal{R}_{d}\}$ \emph{if there exists} $\rho_{d}\in\mathcal{R}_{d}$ for
all $d$, such that
\begin{equation}
\lambda\left(  a\right)  =%
{\displaystyle\sum\limits_{d\in\mathcal{D}}}
Q\left(  d\right)  \rho_{d}(a)\ \ \text{for all }a\in X. \label{Rat}%
\end{equation}

Perfect information is the special case where each $\mathcal{R}_{d}$ is a
singleton. More interesting specifications follow.\footnote{They are all
well-known in both robust statistics and in decision theory. We have borrowed
them and their properties described below from Wasserman and Kadane (1990).
However, we have not seen the last three used previously in the present
context.}

\bigskip

\noindent\textbf{Complete ignorance}: Let $\mathcal{R}_{d}=\Delta\left(
C_{d}\right)  $ for each $d$ as indicated above.

\bigskip

\noindent$\epsilon$-\textbf{contamination}: For each $d$, let $\widehat{\rho
}_{d}\in\Delta\left(  C_{d}\right)  $ be a focal probability distribution over
alternatives, perhaps the analyst's best "point estimate," but one in which
she may not have complete confidence. As a reflection of her incomplete
confidence she entertains as possible all contaminations of $\widehat{\rho
}_{d}$ of the form
\[
\rho_{d}=\left(  1-\epsilon\right)  \widehat{\rho}_{d}+\epsilon\widetilde
{\rho}_{d}\text{,}%
\]
where $\widetilde{\rho}_{d}$ is any measure on $C_{d}$ and where
$0\leq\epsilon\leq1$ is a parameter to be specified by the analyst. That is,
let
\begin{align}
\mathcal{R}_{d}  &  =\{\rho_{d}:\rho_{d}=\left(  1-\epsilon\right)
\widehat{\rho}_{d}+\epsilon\widetilde{\rho}_{d},~\widetilde{\rho}_{d}\in
\Delta\left(  C_{d}\right)  \}\label{R-epsilon}\\
&  =\left(  1-\epsilon\right)  \widehat{\rho}_{d}+\epsilon\Delta\left(
C_{d}\right)  \text{.}\nonumber
\end{align}
The extremes $\epsilon=0,1$ correspond respectively to the complete confidence
and complete ignorance models respectively. Further, it is easy to see that
$\mathcal{R}_{d}$ grows larger in the sense of set inclusion as $\epsilon$
increases in $\left[  0,1\right]  $. This suggests the interpretation of
decreasing confidence (or increasing ignorance) as $\epsilon$ increases.

The "$\epsilon$-contamination" model was proposed initially by Huber (1964)
and has been used frequently in robust statistics (e.g. Huber and Ronchetti
2009, Wasserman and Kadane 1990), and also in decision theory and its many
applications where it is a useful parametric specialization of the set of
priors appearing in multiple-priors utility (Gilboa and Schmeidler 1989).

\bigskip

\noindent\textbf{Variation neighborhood}: For any $p^{\prime}$ and $p$ in
$\Delta\left(  C_{d}\right)  $, define%
\[
\delta_{d}\left(  p^{\prime},p\right)  =\sup_{K\subset C_{d}}\mid p^{\prime
}\left(  K\right)  -p\left(  K\right)  \mid\text{,}%
\]
Fix a reference/focal measure $P_{d}$ on $C_{d}$ and $\epsilon>0$, and let%
\begin{equation}
\mathcal{R}_{d}=\{p_{d}\in\Delta\left(  C_{d}\right)  :\delta_{d}\left(
p_{d},P_{d}\right)  <\epsilon\}\text{.}~ \label{R-nbhd}%
\end{equation}

\bigskip

\noindent\textbf{Interval belief}: Let $p_{d\ast}$ and $p_{d}^{\ast}$ be
measures (not probability measures) on $C_{d}$, satisfying%
\[
p_{d\ast}\left(  \cdot\right)  \leq p_{d}^{\ast}\left(  \cdot\right)  \text{
and }0<p_{d\ast}\left(  C_{d}\right)  <1<p_{d}^{\ast}\left(  C_{d}\right)
\text{,}%
\]
and define
\[
\mathcal{R}_{d}=\{p_{d}\in\Delta\left(  C_{d}\right)  :~p_{d\ast}\left(
\cdot\right)  \leq p_{d}\left(  \cdot\right)  \leq p_{d}^{\ast}\left(
\cdot\right)  \text{ on }C_{d}\}\text{.}%
\]
In the special case
\[
p_{d\ast}=a_{d}P_{d}\text{ and }p_{d}^{\ast}=b_{d}P_{d}\text{,}%
\]
where $a_{d}<1<b_{d}$ and $P_{d}~$is a probability measure on $C_{d}$, one
obtains
\[
\mathcal{R}_{d}=\{p_{d}\in\Delta\left(  C_{d}\right)  :~a_{d}P_{d}\leq
p_{d}\leq b_{d}P_{d}\}
\]

\medskip

In all cases, the identified set is (weakly) smaller than the identified set
under complete ignorance. More generally, it shrinks if confidence increases
in the sense that each set $\mathcal{R}_{d}$ shrinks; this happens, for
example, if $\epsilon$ is reduced in the $\epsilon$-contamination
specification or in the variation neighborhood specification. (Similarly, if
each set of alternatives $C_{d}$ shrinks.) It is easy to see also that an
increase in confidence shrinks the set of empirical measures $\lambda$ that
can be rationalized by some $Q$. For example, in the absence of any confidence
(complete ignorance), every $\lambda$ with support in $\cup_{d}C_{d}$ can be
rationalized by some $Q$, while in the $\epsilon$-contamination specification
$\lambda$ can be rationalized only if it can be expressed as a mixture
$\left(  1-\epsilon\right)  \widehat{\lambda}+\epsilon\widetilde{\lambda}$
where $\widehat{\lambda}$ is rationalizable under complete confidence
($\epsilon=0$) and $\widetilde{\lambda}$ is rationalizable under zero
confidence ($\epsilon=1$), which is equivalent to the statement
\[
\lambda\in\left(  1-\epsilon\right)  ch\left(  \{\widehat{\rho}_{d}\}\right)
+\epsilon\Delta\left(  \cup_{d}C_{d}\right)  \text{,}%
\]
where $ch$ denotes `convex hull.'

\subsection{A characterization\label{section-thm}}

The main question to be addressed is "\textit{which measures }$Q$ \textit{can
rationalize }$\lambda$\textit{ given} $\{\mathcal{R}_{d}\}$?" We provide a
comprehensive answer under the assumption that each $\mathcal{R}_{d}$ is the
core of a convex capacity, that is, for each $d$,\footnote{Since a convex
capacity is uniquely determined by its core (see (\ref{lowerbd})), $\nu_{d}$
is necessarily unique. Another point (see the appendix), is that a capacity
$\nu_{d}$ on $C_{d}$, hence satisfying $\nu_{d}(C_{d})=1$, can be uniquely
extended to a capacity on all of $X$, just as a probability measure on $C_{d}$
can be so extended.}%
\begin{equation}
\mathcal{R}_{d}=core\left(  \nu_{d}\right)  \text{, for some }\nu_{d}\text{
convex.} \label{Ri-core}%
\end{equation}
Though limiting, (\ref{Ri-core}) is of interest in light of the role of convex
capacities in decision theory and in statistics (as mentioned in the
introduction); and we note also that it is satisfied by all of the preceding
specifications (as shown shortly).

\begin{thm}
\label{thm-main}Let $\{\mathcal{R}_{d}\}$ be such that, for each
$d\,$,$\ \mathcal{R}_{d}=core\left(  \nu_{d}\right)  $ for some convex
capacity $\nu_{d}$ on $C_{d}$. Then $Q\in\Delta\left(  \mathcal{D}\right)  $
rationalizes $\lambda$ given $\{\mathcal{R}_{d}\}$ if and only if
\begin{equation}
\lambda\left(  K\right)  \geq%
{\displaystyle\sum\limits_{d\in\mathcal{D}}}
Q\left(  d\right)  \nu_{d}\left(  K\cap C_{d}\right)  \text{ \ for all
}K\subset X\text{.} \label{dominate}%
\end{equation}
In particular, this equivalence applies to the four special cases of
$\{\mathcal{R}_{d}\}$ described above where the corresponding capacities
$\nu_{d}$ are given by:\newline%
\[%
\begin{array}
[c]{cccc}%
ignorance & \nu_{d}\left(  K\right)  & = & \boldsymbol{1}_{C_{d}\subset K}\\
&  &  & \\
contamination: & \nu_{d}\left(  K\right)  & = & \left(  1-\epsilon\right)
\widehat{\rho}_{d}\left(  K\cap C_{d}\right)  +\epsilon\boldsymbol{1}%
_{C_{d}\subset K}\\
&  &  & \\
variation\text{ }nbhd & \nu_{d}\left(  K\right)  & = &
\begin{array}
[c]{c}%
\max\{P_{d}\left(  K\cap C_{d}\right)  -\epsilon,0\}\text{ if }C_{d}%
\not \subset K\\
\text{and }=1\text{ if }C_{d}\subset K
\end{array}
\\
&  &  & \\
interval\text{ }beliefs & \nu_{d}\left(  K\right)  & = &
\begin{array}
[c]{c}%
\max\{p_{d\ast}\left(  K\cap C_{d}\right)  ,p_{d}^{\ast}\left(  K\cap
C_{d}\right)  -\beta_{d}\}\\
\beta_{d}=p_{d}^{\ast}\left(  C_{d}\right)  -1
\end{array}
\end{array}
\]

\end{thm}

\medskip

The main message is that the sharp identified set of measures $Q$ is the set
of solutions $Q$ to the finite set of linear inequalities (\ref{dominate}),
and constitutes a (convex) polytope. The proof is extremely simple.

\medskip

\noindent\textbf{\noindent Proof: }Under the assumption (\ref{Ri-core}),
rationalizability amounts to the statement that
\[
\lambda\in\sum_{d}Q\left(  d\right)  ~core\left(  \nu_{d}\right)  ,
\]
while condition (\ref{dominate}) is the statement that
\[
\lambda\in core\left(  \sum_{d}Q\left(  d\right)  \nu_{d}\right)  \text{.}%
\]
However, by (\ref{mixturelinear}), the core of the mixture equals the mixture
of the cores, which proves the required equivalence.

For the assertions regarding the special cases, one need only show that in
each case the indicated capacity $\nu_{d}$ is convex and that it has core
equal to the corresponding set $\mathcal{R}_{d}$. But these are well-known
facts (Huber and Strassen 1973, Wasserman and Kadane 1990).{\normalsize \hfill
$\blacksquare$}

\medskip

\noindent\textbf{Remark:} It is noteworthy that the proof uses convexity of
the $\nu_{d}$s only to justify applying the mixture-linearity property of
their cores. That is, the characterization provided by (\ref{dominate}) is
valid also if convexity is replaced by this mixture-linearity. In a sense,
therefore, since Strassen's (1965) Theorem 4 assumes convexity, our result is
(strictly) more general. Tijs and Branzei (2002) and Bloch and de Clippel
(2010) give other assumptions, besides convexity, that imply mixture-linearity
of cores.\footnote{They work in the context of cooperative games where
capacities are typically not normalized to assign a common fixed value to the
universal coalition, and hence they refer to additivity rather than
mixture-linearity of the core.} Indeed, it follows from the latter paper that
(generically) there exists a partition of the set of all capacities having
nonempty cores such that the mixture-linearity property is satisfied if (and
only if) all capacities lie in the same equivalence class. The set of convex
capacities is one such equivalence class, but there are others and the theorem
applies to each of them as well. It remains for future work to determine if
any of the other equivalence classes provide alternatives to convexity that
are interesting in our setting.

\medskip\ 

A brief discussion of the ignorance special case may be clarifying and provide
perspective on the theorem. The associated capacities, written more fully, are
given by
\[
\nu_{d}\left(  K\right)  =\left\{
\begin{array}
[c]{ccc}%
1 &  & C_{d}\subset K\\
0 &  & C_{d}\not \subset K
\end{array}
\right.
\]
The epistemic interpretation is that $C_{d}$ is certain but there is complete
ignorance within $C_{d}$. The condition (\ref{dominate}) characterizing
rationalizability specializes to the set of inequalities%
\begin{equation}
\lambda\left(  K\right)  \geq Q\left(  \left\{  d\in\mathcal{D}:C_{d}\subset
K\right\}  \right)  \text{ ~for all }K\subset X\text{.} \label{dominate0}%
\end{equation}
Similar conditions have appeared previously in Barseghyan et al (2021, Theorem
3.1) and in Azrieli and Rehbeck (2023). As indicated in the introduction, the
latter addresses different questions, and the former assumes a designated
minimum size for menus. The ignorance special case admits alternative proofs.
For one, the associated capacities $\nu_{d}$ are belief functions and hence
the result for that case can be derived by using random sets, which is the
approach taken by Barsheghyan et al (2021). Alternatively, it follows
immediately from the well-known structure of the core of a belief function
(Dempster (1967) or Wasserman (1990, Theorem 2.1)). Moreover, these
alternatives apply also to the $\epsilon$-contamination specification since
its $\nu_{d}$s are belief functions. However, they do not apply when $\nu_{d}%
$s are convex but not belief functions, such as in the other two special cases
or at the level of generality in the theorem.

\bigskip

Application of the theorem requires that when considering whether to adopt a
specification of interest for the $\mathcal{R}_{d}$s one is able to check
whether it satisfies (\ref{Ri-core}). For the particular specifications
addressed in the theorem, the literature has confirmed (\ref{Ri-core}). More
generally, an important observation is that, given $\{\mathcal{R}_{d}\}$,
then, for each $d$, there is only one candidate for a suitable capacity
$\nu_{d}$, namely the \emph{lower probability} corresponding to $\mathcal{R}%
_{d}$ and defined by
\[
\nu_{d}\left(  K\right)  =\inf\{\rho\left(  K\right)  :\rho\in\mathcal{R}%
_{d}\}\text{, \ for all }K\subset X\text{.}%
\]
In other words, (\ref{Ri-core}) is \emph{equivalent} to the assumption that
the lower probability capacity is convex and has $\mathcal{R}_{d}$ as its
core. (This follows directly from (\ref{lowerbd}).) Convexity of $\nu_{d}$ can
be checked, in principle, by using its definition (\ref{super}) or any of its
equivalent characterizations (Grabisch (2016, Theorem 3.15), for example).
Since $\mathcal{R}_{d}\subset core\left(  \nu_{d}\right)  $ follows from the
definition of lower probability, equality amounts to the requirement that
$\mathcal{R}_{d}$ be sufficiently large in the sense that, for every $\rho
\in\Delta\left(  C_{d}\right)  $,
\[
\rho\left(  K\right)  \geq\nu_{d}\left(  K\right)  \,\ \text{for all }K\subset
C_{d}\text{ implies }\rho\in\mathcal{R}_{d}\text{.}\footnote{Alternatively,
given convexity, one can compute the cores, for example, by using the greedy
algorithm Ichiishi (1981), or the algorithm in Chambers and Melkonyan (2005)
that uses information about willingness to buy or sell and thus may help the
analyst to calibrate parameters like $\epsilon$.}%
\]

Related is the question what can be done if one drops the assumption
(\ref{Ri-core}) entirely.\footnote{If the sets $\{C_{d}\}$ are disjoint, then,
for any $\{\mathcal{R}_{d}\}$, there is point identification - $\lambda$ is
rationalized by the \emph{unique} measure $Q$ given by $Q\left(  d\right)
=\lambda\left(  C_{d}\right)  $ for all $d$.} In fact, it is straightforward
to show that the counterpart of (\ref{dominate}) given below is
\emph{necessary} for rationalizability given $\{\mathcal{R}_{d}\}$: If
$Q\in\Delta\left(  \mathcal{D}\right)  $ rationalizes $\lambda$, then, for the
lower probability capacity $\nu_{d}$, (using $\mathcal{R}_{d}\subset
core\left(  \nu_{d}\right)  $ and (\ref{mixture0})),
\begin{align*}
\lambda &  \in%
{\displaystyle\sum\limits_{d}}
Q\left(  d\right)  \mathcal{R}_{d}\subset%
{\displaystyle\sum\limits_{d}}
Q\left(  d\right)  core\left(  \nu_{d}\right)  \\
&  \subset core\left(
{\displaystyle\sum\limits_{d}}
Q\left(  d\right)  \nu_{d}\right)  \ \Longrightarrow\\
\lambda\left(  K\right)   &  \geq%
{\displaystyle\sum\limits_{d}}
Q\left(  d\right)  \nu_{d}\left(  K\cap C_{d}\right)  \text{ \ for all
}K\subset X\text{.}%
\end{align*}

\medskip\bigskip

The theorem is relevant also to a rationalizability notion such as
(\ref{Rat0}), suitably modified, where menus appear explicitly. Modify
(\ref{Rat0}) by allowing $\pi_{d}$, for each $d$, to vary over a set $\Pi
_{d}\subset\Delta\left(  \mathcal{A}\right)  $, where $\Pi_{d}$ is determined
by the analyst. (For example, $\Pi_{d}=\Delta\left(  \mathcal{A}\right)  $
would model complete ignorance about menus.) Say that $Q\in\Delta\left(
\mathcal{D}\right)  $ \emph{menu-rationalizes} $\lambda$ \emph{given}
$\{\Pi_{d}\}$ \emph{if there exists} $\pi_{d}\in\Pi_{d}$ for all $d$, such that%

\begin{equation}
\lambda\left(  a\right)  =%
{\displaystyle\sum\limits_{d\in\mathcal{D}}}
{\displaystyle\sum\limits_{A\in\mathcal{A}}}
Q\left(  d\right)  \pi_{d}(A)~\boldsymbol{1}_{d\left(  A\right)  =a},\text{
\ for all }a\in X\text{.} \label{Rat-menu}%
\end{equation}
Each $\pi_{d}$ induces a distribution $\rho_{d}$ on $C_{d}$ as in
(\ref{rhoi}); let $\mathcal{R}_{d}$ be the set of all such distributions
$\rho_{d}$ as $\pi_{d}$ varies over $\Pi_{d}$. Then it is immediate that
menu-rationalization of $\lambda$ given $\{\Pi_{d}\}$ implies rationalization
(\ref{Rat}) given $\{\mathcal{R}_{d}\}$. Moreover, if $\Pi_{d}$ is the core of
a convex capacity on $\mathcal{A}$, then $\mathcal{R}_{d}$ is the core of a
convex capacity $\nu_{d}$ on $C_{d}$ (see appendix). Hence (\ref{dominate}) is
necessary for menu-rationalizability by $Q$. To prove sufficiency, suppose
that $Q$ rationalizes $\lambda$ given $\{\mathcal{R}_{d}=core\left(  \nu
_{d}\right)  \}$ and define $\{\Pi_{d}\}\subset\Delta\left(  \mathcal{A}%
\right)  $ as follows: for each $d$ and $\rho_{d}\in\mathcal{R}_{d}$, and for
each $a\in C_{d}$, select one menu $A_{a,d}$ satisfying $d\left(
A_{a,d}\right)  =a$, and define
\[
\pi_{d}\left(  A\right)  =\left\{
\begin{array}
[c]{ccc}%
\rho_{d}\left(  a\right)  &  & A=A_{a,d}\\
0 &  & A\not =A_{a,d}%
\end{array}
\right.
\]
Then $\pi_{d}$ is a probability measure because
\[
\sum_{A\in\mathcal{A}}\pi_{d}\left(  A\right)  =\sum_{a\in C_{d}}\rho
_{d}\left(  a\right)  =\rho_{d}\left(  C_{d}\right)  =1\text{.}%
\]
Let $\Pi_{d}$ be the set of all such $\pi_{d}$s as $\rho_{d}$ varies over
$\mathcal{R}_{d}$. Then $Q$ menu-rationalizes $\lambda$ given $\{\Pi_{d}\}$
because
\begin{align*}
\pi_{d}\left(  \{A\in\mathcal{A}:d\left(  A\right)  =a\}\right)   &  =\rho
_{d}\left(  a\right)  \Longrightarrow\\%
{\displaystyle\sum\limits_{d}}
Q\left(  d\right)  \sum_{A\in\mathcal{A}}\pi_{d}\left(  A\right)
\boldsymbol{1}_{d\left(  A\right)  =a}  &  =%
{\displaystyle\sum\limits_{d}}
Q\left(  d\right)  \rho_{d}\left(  a\right) \\
&  =\lambda\left(  a\right)  \text{.}%
\end{align*}
(Note that under complete ignorance ($\mathcal{R}_{d}=\Delta\left(
C_{d}\right)  $ or $\Pi_{d}=\Delta\left(  \mathcal{A}\right)  $), the above
proves the equivalence of the two notions of rationalizability.)

\bigskip

The following two examples relate to questions that arise from Barseghyan et
al (2021) and Azrieli and Rehbeck (2023) respectively.

\medskip

\noindent\textbf{Example 1 (singleton menus)}: If $\mathcal{A}$ includes all
singletons and there is complete ignorance about menus, then $C_{d}=X$ for all
$d$ and any $\lambda$ is rationalized by any $Q$. Barseghyan et al (2021)
assume that the minimum menu size is at least two to avoid this scenario in
their setup. However, in our setup we can allow a subset of all singleton
menus, in which case they can affect (strictly expand) the identified region.

We illustrate here for the case where every decision rule $d$ is derived from
maximization of a preference order, and where complete ignorance is assumed.
Then each $Q$ describes a probability distribution over preferences. Let
$X=\{a,b,c\}$. The six possible preference orders are:
\begin{align*}
a  &  \succ_{1}b\succ_{1}c,a\succ_{2}c\succ_{2}b\\
b  &  \succ_{3}a\succ_{3}c,b\succ_{4}c\succ_{4}a\\
c  &  \succ_{5}a\succ_{5}b,c\succ_{6}b\succ_{6}a\text{.}%
\end{align*}
Finally, let $\mathcal{A}=\{\{a\},\{a,b\},\{a,b,c\}\}$.\footnote{Thus menus
are nested. Think, for example, of expanding budget sets.} Then, after
deleting redundant inequalities, (\ref{dominate0}) reduces to:
\begin{align*}
\lambda\left(  a\right)   &  \geq Q\left(  \succ_{1}\right)  +Q\left(
\succ_{2}\right) \\
\lambda\left(  \{a,b\}\right)   &  \geq Q\left(  \succ_{1}\right)  +Q\left(
\succ_{2}\right)  +Q\left(  \succ_{3}\right)  +Q\left(  \succ_{4}\right) \\
\lambda\left(  \{a,c\}\right)   &  \geq Q\left(  \succ_{1}\right)  +Q\left(
\succ_{2}\right)  +Q\left(  \succ_{5}\right)
\end{align*}
The first inequality provides an upper bound on the probability of the set of
preferences that rank alternative $a$ highest, the second provides an upper
bound on the probability of preferences that rank $a$ or $b$ highest, and the
third gives an upper bound on the probability of preferences that rank $a$
above $b$.

Consider now what happens if the singleton $\{a\}$ is deleted from
$\mathcal{A}$. Let $\mathcal{A^{\prime}}=\{\{a,b\},\{a,b,c\}\}$ be the new set
of menus. Then the following \emph{additional} inequalities are implied by
(\ref{dominate0}):
\begin{align*}
\lambda\left(  b\right)   &  \geq Q\left(  \succ_{3}\right)  +Q\left(
\succ_{4}\right)  \\
\lambda\left(  \{b,c\}\right)   &  \geq Q\left(  \succ_{3}\right)  +Q\left(
\succ_{4}\right)  +Q\left(  \succ_{6}\right)
\end{align*}
As a result, the sharp identification region shrinks strictly. (The intuition
is that when $\{a\}$ is removed, then the sets $C_{3}$ and $C_{4}$ shrink,
leading to the lower bound for $\lambda\left(  b\right)  $, and $C_{6}$
shrinks, which leads to the lower bound for $\lambda\left(  \{b,c\}\right)  $.)

For a numerical example, take $\lambda\left(  a\right)  =1/2,\lambda\left(
b\right)  =1/4,$ and $\lambda\left(  c\right)  =1/4$. The preference
distribution given by $Q\left(  \succ_{1}\right)  =Q\left(  \succ_{3}\right)
=Q\left(  \succ_{4}\right)  =Q\left(  \succ_{6}\right)  =1/4$ rationalizes
$\lambda$ when $\{a\}$ is included, but not if it is removed. The presence of
a singleton menu does not preclude meaningful inference, but it does weaken
inference by expanding the sharp identification region.

\medskip

\noindent\textbf{Example 2 (menu homogeneity)}: Menu-rationalizability as
defined in (\ref{Rat-menu}) permits heterogeneity in both decision rules
\emph{and} in the menu formation processes, the latter because $\pi_{d}$ and
$\pi_{d^{\prime}}$ are allowed to differ. Refer to \emph{menu-homogeneity} if
$\pi_{d}=\pi$ for all $d$.
This hypothesis has been adopted in several applied works where one can
interpret the different menus as arising from feasibility rather than
consideration (Tenn and Yun 2008, Tenn 2009, Conlon and Mortimer 2013, Lu
2022), and in the theoretical contribution by Azrieli and Rehbeck (2023,
section 4), while its limitations have been noted by Barsheghyan et al
(2021).\textbf{ }Where menus are based on consideration one would expect them
to depend on preference (or decision rule), as in the applied papers by Goeree
(2008), and Abaluck and Adams-Prassl (2021). Here we demonstrate that imposing
menu-homogeneity\ can lead to different conclusions about the sharp identified set.

Let $X=\{a,b,c\}$, and assume preference maximization, with only two possible
preference orders
\[
a\succ_{1}b\succ_{1}c\text{, \ }a\succ_{2}c\succ_{2}b\text{.}%
\]
Finally, let $\mathcal{A}=\{\{a,b\},\{a,c\},\{b,c\},\{a,b,c\}\}$, and let the
empirical measure be given by
\[
\lambda\left(  a\right)  =\lambda\left(  b\right)  =\lambda\left(  c\right)
=1/3.
\]

\noindent Then $\lambda$ can be rationalized by $Q$, where $Q\left(  \succ
_{1}\right)  =2/3$ and $Q\left(  \succ_{2}\right)  =1/3$, because the
inequalities (\ref{dominate0}) can be verified. A corresponding version of
(\ref{Rat-menu}) uses the (heterogeneous) distributions over menus given by
\[
\pi_{1}\left(  \{b,c\}\right)  =1/2\;\text{and }\pi_{2}\left(  \{b,c\}\right)
=1\text{.}%
\]
However, $Q$ cannot rationalize $\lambda$ if one insists on menu-homogeneity:
Under the latter condition, (\ref{Rat-menu}) implies
\begin{align*}
\lambda\left(  b\right)   &  =Q(\succ_{1})[\pi\left(  \{b,c\}\right)  ]=1/3\\
\lambda\left(  c\right)   &  =Q(\succ_{2})[\pi\left(  \{b,c\}\right)
]=1/3\text{,}%
\end{align*}
which would force $Q$ to assign equal probabilities to both preferences, a contradiction.

In general, any restriction on the distributions over menus admitted in
(\ref{Rat-menu}) makes menu-rationalizability more difficult and thus shrinks
the sharp identified set. The example confirms that in the case of
menu-homogeneity the shrinkage can be strict. (Finally, note that in the
example the sharp identification set with menu-homogeneity is not empty. For
the two preference orders given, the unique rationalizing measure\textbf{ }$Q$
assigns equal probability to the two preference orders.)

\section{Identifying updating rules \label{section-updating}}

Given the motivation described in the introduction, proceed as follows. Let
$\Omega=\{\omega_{a},\omega_{b}\}$ be the set of possible future states of the
world. (The binary specification is enough to capture uncertainty about
whether an event will or will not happen.) There is a common prior
$\mu=\left(  \mu_{a},\mu_{b}\right)  >0$. Every individual $i$ receives some
information, (or observes a signal), and updates beliefs to the posterior
$p_{i}=\left(  p_{ia},p_{ib}\right)  >0$. Beliefs are represented equivalently
by their implied \emph{odds ratios} (in logarithmic form)
\[
x^{0}=\log\left(  \mu_{a}/\mu_{b}\right)  \text{ and }x_{i}=\log(p_{ia}%
/p_{ib})\newline\text{.}%
\]
These ratios lie in a finite set $X$.\footnote{One can specify primitives of
the model so that sets below are finite wherever needed by the formalism. We
will not elaborate.} The data to be rationalized is $\lambda\in\Delta\left(
X\right)  $, the empirical cross-sectional distribution of posterior odds
ratios. Neither the updating rules used by individuals nor their information
(viewed as being generated by an `experiment') is observable to the analyst.

Our formal model begins with the representation of experiments. Since a signal
leads to a posterior odds ratio $x$ in $X$ and hence to the change from the
prior view equal to $x-x^{0}$, we define an $\emph{experiment}$ to be a
probability measure $E$ in $\Delta\left(  X-\{x^{0}\}\right)  $%
.\footnote{$X-\{x^{0}\}=\{x-x^{0}:x\in X\}$.} To motivate this definition
further, note that in the Bayesian framework standard application of Bayes
rule implies%
\begin{equation}
x=LLR+x^{0}\text{,} \label{xLLR}%
\end{equation}
where $LLR$ denotes the \emph{log- likelihood ratio}. Thus every experiment
amounts to a distribution over log-likelihood ratios. (Equivalently, it is a
distribution over posteriors odds ratio $x$, or over the posterior probability
measure which is how experiments are frequently described.) Though the
preceding presumes Bayesian updating, we show shortly that (\ref{xLLR}) is
compatible also with other updating rules. Accordingly, we continue to refer
to signals as LLRs and as points in $X-\{x^{0}\}$.

The set of all experiments deemed relevant by the analyst, and
satisfying\textbf{ }$E\left(  0\right)  <1$, is $\mathcal{E}$. An
\emph{updating rule} $\psi$ is a mapping from an experiment into a
distribution over posterior odds:
\[
\psi\left(  E\right)  \in\Delta\left(  X\right)  \text{.}%
\]
Based on (\ref{xLLR}), the focal \emph{Bayesian updating rule} $\psi^{BR}$ is
given by \textbf{ }
\begin{equation}
\left(  \psi^{BR}\left(  E\right)  \right)  \left(  x\right)  =E\left(
x-x^{0}\right)  \text{, \ for all }x\in X\text{,} \label{BR}%
\end{equation}
for all $E\in\mathcal{E}$, that is, under Bayesian updating, the change in
odds $x-x^{0}$ has the same distribution as $LLR$.

We also admit non-Bayesian updating rules defined as follows: For each
parameter $\kappa$, \underline{$\kappa$}$\leq\kappa\leq1$, $\psi^{\kappa}$ is
given by\footnote{$\delta_{x^{0}}$ is the measure on $X$ that assigns
probability 1 to the prior odds ratio $x^{0}$.}%
\begin{equation}
\psi^{\kappa}\left(  E\right)  =\left(  1-\kappa\right)  \psi^{BR}\left(
E\right)  +\kappa\delta_{x^{0}}\text{,} \label{psi-k}%
\end{equation}
where the lower bound \underline{$\kappa$} satisfies
\[
\underline{\kappa}\geq-E\left(  0\right)  /\left(  1-E\left(  0\right)
\right)  \text{ for all }E\text{.}%
\]
(The latter condition implies that, for all $E$, $\psi^{\kappa}\left(
E\right)  \left(  x\right)  \geq0$ for all $x$, and hence that $\psi^{\kappa
}\left(  E\right)  \left(  \cdot\right)  $ is a probability measure on $X$.)
For instance, if we take Bayesian updating as the "correct" balancing of prior
beliefs and responsiveness to signals, then, because $\delta_{x^{0}}$ gives no
weight to signals, $\kappa>0$ implies "too much" weight to prior beliefs and
thus \emph{underreaction} to information. In contrast, $\kappa<0$ is
interpretable as implying \emph{overreaction}. The negative weight given to
the prior view represented by $\delta_{x^{0}}$ upsets the proper balance
captured by $\psi^{BR}\left(  E\right)  $ and implies "too much" sensitivity
to the signal.\footnote{Epstein (2006) axiomatizes a closely related model of
updating, (formulated for absolute probabilities rather than odds ratios), and
we adapt his interpretations to (\ref{psi-k}). Augenblick, Lazarus and Thaler
(2023) postulate updating rules in terms of odds ratios and study how
under/over-reaction varies with the strength of the signal as measured by LLR.
They consider data on individual updating in response to a known experiment.}

Let $\Psi$ be a (finite) subset of $\{\psi^{\kappa}:\underline{\kappa}%
\leq\kappa\leq1\}$ that includes $\psi^{0}=\psi^{BR}$, and, for each $\psi
\in\Psi$, let
\[
\mathcal{P}_{\psi}=\{\psi\left(  E\right)  :E\in\mathcal{E}\}\text{,}%
\]
the set of all distributions over odds ratios induced by an experiment from
$\mathcal{E}$ and an update rule from $\Psi$. The analyst assumes only that
every individual employs an experiment $E$ from $\mathcal{E}$ and an updating
rule $\psi$ from $\Psi$, (their distribution across individuals is unknown),
in which case she faces the set $\mathcal{P}_{\psi}$ of posterior
distributions over odds ratios. Roughly, the foursome $\left(  X,\Psi
,\mathcal{E},\{\mathcal{P}_{\psi}\}\right)  $ is the counterpart of the
foursome $\left(  X,\mathcal{D},\mathcal{A},\{\mathcal{R}_{d}\}\right)  $ in
the main choice model.

As the counterpart of (\ref{Rat}) for this setting, say that $Q\in
\Delta\left(  \Psi\right)  $ \emph{rationalizes }$\lambda$ \emph{given}
$\{\mathcal{P}_{\psi}\}_{\psi\in\Psi}$ if, for every $\psi$, there exists
$\rho_{\psi}\in\mathcal{P}_{\psi}$ such that%
\begin{equation}
\lambda\left(  x\right)  =%
{\displaystyle\sum\limits_{\psi\in\Psi}}
Q\left(  \psi\right)  \rho_{\psi}(x)\ \ \text{for all }x\in X.
\label{Rat-update}%
\end{equation}

\noindent The sum is over distinct update rules and thus $Q$ describes their
distribution explicitly.

The conditions characterizing rationalizability follow from Theorem
\ref{thm-main} if the sets $\mathcal{P}_{\psi}$ satisfy the appropriate form
of the core condition (\ref{Ri-core}).\footnote{The theorem is applicable also
given any other specifications of $\mathcal{E}$ and $\Psi$ for which the
implied sets $\mathcal{P}_{\psi}$ satisfy the noted condition. In principle,
therefore, alternative non-Bayesian updating rules could be accommodated.}
Rather than simply assuming the latter, we derive it from the following
assumption about the set of experiments $\mathcal{E}\subset\Delta\left(
X-\{x^{0}\}\right)  $: There exists a convex capacity $\nu$ on $X-\{x^{0}\}$
such that
\begin{equation}
\mathcal{E}=core\left(  \nu\right)  \text{.}\label{core-E}%
\end{equation}
Section \ref{section-main} describes and motivates examples of convex
capacities and their cores that could be used to specify $\mathcal{E}$ and
$\nu$ satisfying (\ref{core-E}). In addition, (\ref{core-E}) implies that, for
each $\psi^{\kappa}\in\Psi$,%

\begin{equation}
\mathcal{P}_{\psi^{\kappa}}=core\left(  \nu^{\kappa}\right)  \label{core-P}%
\end{equation}
for some convex capacity on $X$ (in particular, $\mathcal{P}_{\psi^{\kappa}}$
is a convex set). In fact, for every $K\subset X$,\footnote{When
$K=\varnothing$, take $K-\{x^{0}\}=\varnothing$ and\ hence $\nu\left(
K-\{x^{0}\}\right)  =0$.} \
\begin{equation}
\nu^{\kappa}\left(  K\right)  =(1-\kappa)\nu\left(  K-\{x^{0}\}\right)
+\kappa\delta_{x^{0}}\left(  K\right)  \text{.} \label{vi}%
\end{equation}

Conclude from Theorem \ref{thm-main} that $Q$ rationalizes $\lambda$ in the
sense of (\ref{Rat-update}) if and only if, for all $K\subset X$,
\begin{equation}
\lambda\left(  K\right)  \geq\left\{
\begin{array}
[c]{cc}%
(1-\kappa_{Q})\nu\left(  K-\{x^{0}\}\right)  & \text{if }x^{0}\not \in K\\
(1-\kappa_{Q})\nu\left(  K-\{x^{0}\}\right)  +\kappa_{Q} & \text{if }x^{0}\in
K\text{,}%
\end{array}
\right.  \label{lamda-K}%
\end{equation}
where $\kappa_{Q}\equiv\Sigma_{\psi\in\Psi}Q\left(  \psi^{\kappa}\right)
\kappa$ is the \emph{average updating bias} implied by $Q$. Notably, whether
or not $Q$ rationalizes $\lambda$ depends only on the average bias that it
implies. In particular, any rationalization of $\lambda$ can be mimiced by a
population where everyone exhibits this average updating bias. This suggests
the reframing whereby we ask if the average bias $\kappa_{av}$, $\min
_{i}\kappa_{i}\leq\kappa_{av}\leq\max_{i}\kappa_{i}$, can rationalize
$\lambda$.

The following additional implications are straightforward (use (\ref{lamda-K}%
)). There exists an Bayesian average rationalization if and only if, for all
$K\subset X$,
\begin{align*}
\lambda\left(  K\right)   &  \geq\nu\left(  K-\{x^{0}\}\right)  =\inf
_{E\in\mathcal{E}}E\left(  K-\{x^{0}\}\right)  \text{ }\\
&  =\inf_{E\in\mathcal{E}}\psi^{BR}\left(  E\right)  \left(  K\right)
\text{,}%
\end{align*}
where we use (\ref{lowerbd}), that is, $\nu$ is the lower envelope of the set
$\mathcal{E}$ of experiments. \ A non-zero average updating bias is required
to rationalize $\lambda$ under the following conditions. If $\kappa_{av}$ is
any average bias that rationalizes $\lambda$ and if there exists $K^{\ast
}\subset X$ such that
\begin{equation}
\lambda\left(  K^{\ast}\right)  <\nu\left(  K^{\ast}-\{x^{0}\}\right)
=\inf_{E\in\mathcal{E}}\psi^{BR}\left(  E\right)  \left(  K^{\ast}\right)
\text{,} \label{K*}%
\end{equation}
then
\[%
\begin{array}
[c]{cccc}%
\kappa_{av}>0 & \text{if} & x^{0}\not \in K^{\ast} & \text{and}\\
\kappa_{av}<0 & \text{if} & x^{0}\in K^{\ast}\text{,} &
\end{array}
\]
that is, there is \emph{underreaction on average} if $x^{0}\not \in K^{\ast}$
and \emph{overreaction on average} otherwise.

Finally, it follows that $\lambda$ \emph{cannot be rationalized by any average
bias (or measure} $Q$) if there exist subsets $K^{1}$ and $K^{2}$ of $X$ such
that (\ref{K*}) is satisfied by both $K^{\ast}=K^{i}$, $i=1,2$, and if
$x^{0}\not \in K^{1}$, $x^{0}\in K^{2}$.

\section{Concluding illustrations of scope\label{section-conclude}}

We conclude by describing two other settings where our theorem can be applied
to characterize the robust identification of heterogeneity. The first concerns
the identification of (unobservable) heterogeneous effort; notably, the story
is not directly connected to choice or beliefs. The second is an instance of
our choice model where decision rules are not based on preference
maximization; rather, all choices are from a single known objective feasible
set and are due to satisficing with heterogeneous unobserved reservation
values which distribution is identified.

\subsection{Identifying effort}

Consider a population of workers with common observable characteristics (e.g.
education and experience), and working independently. Each produces a
homogeneous output in quantity represented by an element of $X$.\footnote{To
make clear the connection to the main choice model, we use the same symbols,
though with different interpretations.} The empirical frequency distribution
of outputs is given by $\lambda\in\Delta\left(  X\right)  $. Heterogeneity in
output is attributed to differences in unobservable characteristics. The first
unobservable is effort - there are finitely many effort levels $d\in
\mathcal{D}$. The other unobservable is "everything else." The analyst may not
be able to describe these other factors precisely, or even at all. However,
she takes a stand on the set of their possible output consequences. Formally,
for each effort $d$, denote by $C_{d}\subset X$ the set of outputs possible
given the effort level and given what may ensue from "everything else." The
analyst specifies the sets $C_{d}$, but is ignorant about likelihoods
$\emph{within}$ these sets.\footnote{The assumption that $C_{d}$ can be
specified even though "everything else" is poorly understood brings to mind
Maskin and Tirole (1999) who argue that optimal contracts survive even with
unforeseen \emph{contingencies} when agents can forecast future \emph{payoffs}%
.}

With this reinterpretation, rationalizability of $\lambda$ is well-defined,
and Theorem \ref{thm-main} can be applied to yield the (computationally
tractable) sharp identified set of measures $Q$ over effort levels.

The rationalizability notion (\ref{Rat-menu}) could also be accommodated by
introducing a parameter $\theta\in\Theta$ to represent "everything else," and,
for each $d$, a production function $f$ such that the pair $\left(
d,\theta\right)  $ yields output $f\left(  d,\theta\right)  $, and
$C_{d}=\{f\left(  d,\theta\right)  :\theta\in\Theta\}$. Ignorance about
$\Theta$ would be captured by admitting any distribution $\pi_{d}$ over
$\Theta$ in the counterpart of definition (\ref{Rat-menu}). (Roughly, $\Theta$
would play the role of the set of menus $\mathcal{A}$ above.) However, such a
formulation involving production functions $f$ and probability distributions
over $\Theta$ is arguably problematic in situations where the analyst cannot
even conceive of what is included in "everything else."

\subsection{Satisficing}

There is a population of satisficing decision makers whose aspiration
thresholds may differ.\textbf{ }They each choose an alternative from the set
$X$ and they agree that the value of alternatives is described by
$v:X\mapsto\mathbb{R}$. However, individuals differ in two respects. First,
aspiration thresholds differ; the set of distinct thresholds is $\mathcal{V}$.
Second, individuals differ in the order in which they consider alternatives
(this may be a subjective choice of procedure or exogenously imposed). Each
sequential procedure follows a strict total order $>$ on $X$: the individual
chooses the $>$-first alternative with a value at least as large as her
threshold $v\in\mathcal{V}$, and if there are no such "satisfactory"
alternatives then she chooses the $>$-last element in $X$. The empirical
frequency of choices $\lambda$ is observed, but both aspiration levels and
orders $>$ are unobserved. Theorem \ref{thm-main}, suitably reinterpreted, can
be used to partially identify the distribution of aspiration levels while
respecting limited knowledge (or complete ignorance) of the distribution of
orders $>$.

Similar applications can be made to other problems of choice with frames
(Salant and Rubinstein 2008) where frames vary across individuals and are
unobserved by the analyst.

\appendix

\section{Appendix: Basic facts about capacities}

For any finite set $X$, $\nu$ is a \emph{capacity} on $X$ if $\nu
:2^{X}\rightarrow\left[  0,1\right]  $, $\nu\left(  \varnothing\right)  =0$,
$\nu\left(  X\right)  =1$ and $\nu(K^{\prime})\geq\nu(K)$ whenever $K^{\prime
}$ is a superset of $K$. $\nu$ is \emph{convex} if, for all subsets
$K^{\prime}$ and $K$,
\begin{equation}
\nu(K^{\prime}\cup K)+\nu(K^{\prime}\cap K)\geq\nu(K^{\prime})+\nu(K)\text{.}
\label{super}%
\end{equation}
$\nu$ is a \emph{belief function} if, for all $n$, and for all subsets
$K_{1},...,K_{n}$,
\begin{equation}
\nu\left(  \cup_{j=1}^{n}K_{j}\right)  \geq%
{\textstyle\sum\limits_{\varnothing\neq J\subset\left\{  1,...,n\right\}  }}
\left(  -1\right)  ^{\left\vert J\right\vert +1}\nu\left(  \cap_{j\in J}%
K_{j}\right)  \text{.} \label{infinite}%
\end{equation}
If one restricts $n$ to be $2$, then one obtains the condition defining
convexity. Hence every belief function is convex. (Convexity is sometimes
referred to as monotonicity of order 2 while (\ref{infinite}) is called
infinite or total monotonicity.) A more transparent and equivalent definition
of a belief function is that $\nu$ is induced by a random
set.\footnote{Dempster (1967) and Nguyen (1978) are two early references
describing the connection of random sets to belief functions. See also Nguyen
(2006).}

Let $C$ be a subset of $X$ and $\nu$ a capacity on $C$ (hence $\nu\left(
C\right)  =1$). Then $\nu$ can be viewed also as a capacity on $X$ by
identifying $\nu$ with the capacity $\nu^{\prime}$ on $X$ defined by%
\[
\nu^{\prime}\left(  K\right)  =\nu\left(  K\cap C\right)  \text{ for all
}K\subset X\text{.}%
\]
Further, $\nu^{\prime}$ is convex if and only if $\nu$ is convex. We often
identify $\nu$ and $\nu^{\prime}$ and do not distinguish them notationally.

For any capacity $\nu$ on $X$, its core is the set of all dominating
probability measures, that is,
\[
core\left(  \nu\right)  =\left\{  p\in\Delta\left(  X\right)  :p\left(
K\right)  \geq\nu\left(  K\right)  \text{ for all }K\subset X\right\}
\text{.}%
\]
If $\nu$ is convex, then its core is nonempty and $\nu$ can be recovered from
its core as its lower bound or envelope:%
\begin{equation}
\nu\left(  K\right)  =\min\{p(K):p\in core\left(  \nu\right)  \}\text{.}
\label{lowerbd}%
\end{equation}
If $\nu=p$ is a probability measure, then it is convex and $core\left(
\nu\right)  =\{p\}$.

If $\nu$ and $\nu^{\prime}$ are two convex capacities on $X$, and if
$0\leq\alpha\leq1$, then the mixture $\alpha\nu+\left(  1-\alpha\right)
\nu^{\prime}$ is also a convex capacity and its core satisfies
\begin{equation}
core\left(  \alpha\nu+\left(  1-\alpha\right)  \nu^{\prime}\right)  =\alpha
core\left(  \nu\right)  +\left(  1-\alpha\right)  core\left(  \nu^{\prime
}\right)  \text{.} \label{mixturelinear}%
\end{equation}
This "mixture linearity" of the core is the key property that we exploit to
prove our theorem. Elsewhere, we also make use of the following weaker
property that applies to any (not necessarily convex) capacities%
\begin{equation}
core\left(  \alpha\nu+\left(  1-\alpha\right)  \nu^{\prime}\right)
\supset\alpha core\left(  \nu\right)  +\left(  1-\alpha\right)  core\left(
\nu^{\prime}\right)  \text{.} \label{mixture0}%
\end{equation}
See Grabisch (2016, p. 156) for both cases.

Let $\psi$ be a convex capacity on $\mathcal{A}$, $\Pi=core\left(
\psi\right)  $ and $d:\mathcal{A}\longrightarrow X$. Define the (convex) set
$\mathcal{R}$ of all measures $\rho_{\pi}\in\Delta(X)$, where%
\[
\rho_{\pi}\left(  K\right)  =\pi\left(  d^{-1}\left(  K\right)  \right)
\text{, \ for all }K\subset X\text{,}%
\]
and define the set function $\nu$ on $X$ by%
\[
\nu\left(  K\right)  =\psi\left(  d^{-1}\left(  K\right)  \right)  \text{,
\ for all }K\subset X\text{.}%
\]
Then $\nu$ is a convex capacity and $\mathcal{R}=core\left(  \nu\right)  $.
(Convexity follows from verifying (\ref{super}), and $\mathcal{R}\subset
core\left(  \nu\right)  $ is immediate. Let $\{K_{j}\}$ be any chain of
subsets of $X$. Then $\{d^{-1}(K_{j})\}$ is a chain in $\mathcal{A}$. Since
$\psi$ is convex, there exists $\pi^{\ast}\in core\left(  \psi\right)  =\Pi$
such that $\pi^{\ast}\left(  d^{-1}(K_{j})\right)  =\psi\left(  d^{-1}\left(
K_{j}\right)  \right)  $, \ for all $j$ (Choquet 1953). Thus $\rho_{\pi^{\ast
}}\left(  K_{j}\right)  =\nu\left(  K_{j}\right)  $ for all $j$. Apply
Grabisch (2016, Theorem 3.15) to conclude that $\mathcal{R}=core\left(
\nu\right)  $.)


\newpage

\begin{thebibliography}{99}                                                                                               %


\bibitem {Abaluck}Abaluck J. and Adams-Prassl A. What do consumers consider
before they choose? identification from asymmetric demand responses.
\textit{Quart. J. Econ.} (2021) 136, 1611-1663.

\bibitem {Augenblick2023}Augenblick N., Lazarus E. and Thaler M. Overinference
from weak signals and underinference from strong signals. Mimeo 2023.

\bibitem {Rehbeck}Azrieli Y. and Rehbeck J. Marginal stochastic choice. Mimeo 2023.

\bibitem {Ba2023}Ba C., Bohren J.A. and Imas A. Over-and underreaction to
information. Mimeo 2023.

\bibitem {Molinari}Barseghyan L., Coughlin M., Molinari F., and Teitelbaum
J.C. Heterogenous choice sets and preferences. \textit{Econometrica} (2021)
89(5), 2015-2048.

\bibitem {Benjamin2019}Benajmin D.J. Errors in probabilistic reasoning and
judgement biases, pp. 69-186 in \textit{Handbook of Behavioral Economics:
Applications and Foundations 1} 2,\textit{ }D. Bernheim, S. Della Vigna and D.
Laibson eds.\textit{ 2019.}

\bibitem {Bloch2010}Bloch F. and de Clippel G. Cores of combined games.
\textit{J. Econ. Theory} (2010) 145, 2424-2434.

\bibitem {Camerer1995}Camerer C. Individual decision-making, pp. 587-704 in
\textit{The Handbook of Experimental Economics 1}, J. Kagel and A. Roth eds.
Princeton U. Press, 1995.

\bibitem {Caplin}Caplin A., Dean M., and Leahy J. Rational inattention,
optimal consideration sets, and stochastic choice. \textit{Rev. Econ. Stud.}
(2019) 86, 1061-1094.

\bibitem {Cattaneo}Cattaneo M.D., Ma X., Masatlioglu Y., and Suleymanov E. A
random attention model. \textit{J. Pol. Econ.} (2020) 128, 2796-2836.

\bibitem {Chambers2005}Chambers R.G. and Melkonyan T.A. Eliciting the core of
a supermodular capacity. \textit{Econ. Theory} (2005) 26, 203-209.

\bibitem {Choquet1953}Choquet G. Theory of capacities. \textit{Annales de
l'institut Fourier} (1954) 5, 131-295.

\bibitem {Conlon}Conlon C.T. and Mortimer J.H. Demand estimation under
incomplete product availability. \textit{AEJ: Micro} (2013) 5, 1-30.

\bibitem {Dardanoni2020}Dardanoni V., Manzini P., Mariotti M. and Tyson C.J.
Inferring cognitive heterogeneity from aggregate choices.
\textit{Econometrica} (2020) 88(3), 1269-1296.

\bibitem {Dempster1968}Dempster A.P. A generalization of Bayesian inference.
\textit{J. Royal Statist. Soc.} (B) (1968) 30, 205-247.

\bibitem {Dempster1967}Dempster A.P. Upper and lower probabilities induced by
multivalued mappings. \textit{Ann. Math. Statist}. (1967) 38, 325-335.

\bibitem {Doval2023}Doval L. and Eilat R. The core of Bayesian persuasion.
arxiv 2023.

\bibitem {Epstein2006}Epstein L.G. An axiomatic model of non-Bayesian
updating. \textit{Rev. Econ. Stud}. (2006) 73, 413-436.

\bibitem {Filiz-Ozbay}Filiz-Ozbay E. and Masatlioglu Y. Progressive random
choice. \textit{J. Pol. Econ.} (2023) 131, 716-750.

\bibitem {Goeree}Goeree M.S. Limited information and advertising in the U.S.
personal computer industry. \textit{Econometrica} (2008) 76, 1017-1074.

\bibitem {Grabisch}Grabisch M. \textit{Set Functions, Games and Capacities in
Decision Making}. Springer, 2016.

\bibitem {Huber1964}Huber P.J. Robust estimation of a location parameter.
\textit{Ann. Math. Statist}. (1964) 35(1), 73-101.

\bibitem {Huber1981}Huber P.J. and Ronchetti E.M. \textit{Robust Statistics}.
2nd ed. Wiley, New York, 2009.

\bibitem {Strassen1973}Huber P.J. and Strassen V. Minimax tests and the
Neyman-Pearson lemma for capacities. \textit{Ann. Statist}. (1973) 1, 251-263.

\bibitem {Ichiishi1981}Ichiishi T. Super-modularity: applications to convex
games and to the greedy algorithm for LP. \textit{J. Econ. Theory} (1981) 25, 283-286.

\bibitem {Lu}Lu Z. Estimating multinomial choice models with unobserved choice
sets. \textit{J. Econometrics} (2022) 226, 368-398.

\bibitem {Manski1977}Manski C. The structure of random utility models.
\textit{Theory and Decision} (1977) 8(3), 229-254.

\bibitem {Manzini}Manzini P. and Mariotti M. Stochastic choice and
consideration sets. \textit{Econometrica} (2014) 82, 1153-1176.

\bibitem {Masatlioglu}Masatlioglu Y., Nakajima D., and Ozbay E. Revealed
attention. \textit{Amer. Econ. Rev}. (2012) 102, 2183-2205.

\bibitem {Maskin}Maskin E. and Tirole J. Unforeseen contingencies and
incomplete contracts. \textit{Rev. Econ. Stud}. (1999) 66, 83-114.

\bibitem {McFadden}McFadden, D.L. Conditional logit analysis of qualitative
choice behavior, pp. 105-142 in \textit{Frontiers of Econometrics, }ed. P.
Zarembka. New York: Academic Press 1974.

\bibitem {Nguyen1978}Nguyen H.T. On random sets and belief functions.
\textit{J. Math. Anal. Appl.} (1978) 68, 531-542.

\bibitem {Rabin1998}Rabin M. Psychology and economics. \textit{J. Econ. Lit.}
(1998) 36, 11-46.

\bibitem {Salant2008}Salant Y. and Rubinstein A. (A,f): Choice with frames.
\textit{Rev. Econ. Stud.} (2008) 75(4), 1287-1296.

\bibitem {Schmeidler1989}Schmeidler D. Subjective probability and expected
utility without additivity. \textit{Econometrica} (1989) 57(3), 571-587.

\bibitem {Shafer1976}Shafer G. \textit{Mathematical Theory of Evidence}.
Princeton U. Press, 1976.

\bibitem {Strassen1965}Strassen V. The existence of probability measures with
given marginals. \textit{Ann. Math. Statist}. (1965) 36, 81-95.

\bibitem {Tenn2009}Tenn S. Demand estimation under limited product
availability. \textit{Appl. Econ. Letters} (2009) 16, 465-468.

\bibitem {TennYun2008}Tenn S. and Yun J. Biases in demand analysis due to
variation in retail distribution. \textit{Int. J. Ind. Organ.} (2008) 26, 984-997.

\bibitem {Tijs2002}Tijs S. and Branzei R. Additive stable solutions on perfect
cones of cooperative games. \textit{Int. J. Game Theory} (2002) 31, 469-474.

\bibitem {Tversky1974}Tversky A. and Kahneman D. Judgement under uncertainty:
heuristics and biases. \textit{Science} (1974) 185, 1124-1131.

\bibitem {Wasserman1990}Wasserman L.A. Prior envelopes based on belief
functions. \textit{Ann. Statist}. (1990) 18(1), 454-464.

\bibitem {WassermanKadane}Wasserman L.A. and Kadane J.B. Bayes' theorem for
Choquet capacities. \textit{Ann. Math. Statist}. (1990) 18, 1328-1339.
\end{thebibliography}
\end{document}